\documentclass{ws-procs975x65}
\begin{document}
\title{DYNAMICAL FRICTION IN CUSPIDAL GALAXIES}
\author{M. ARCA--SEDDA$^1$ and R. CAPUZZO--DOLCETTA$^2$}

\address{Dep. of Physics, "Sapienza", University of Roma,\\
Roma, 00185, Italy\\
$^1$e-mail: manuel.arcasedda@uniroma1.it\\
$^2$e-mail: roberto.capuzzodolcetta@uniroma1.it\\
}

\begin{abstract}
Dynamical friction is the process responsible for matter transport toward the inner region of galaxies in form of massive objects, like intermediate mass black holes, globular clusters and small satellite galaxies. While very bright galaxies show an almost flat luminosity profile in the inner region, fainter ones have, usually, a peaked, cuspidal, profile toward the center. This makes unreliable, in these cases, the use of the classic Chandrasekhar's formula for dynamical friction in its local approximation. Using both $N$--body simulations and a semi analytical approach, we have obtained reliable results for the orbital decay of massive objects in cuspidal galaxies. 
A relevant result is that of a shallower dependence of dynamical friction braking on the satellite mass than in the usual Chandrasekhar's local expression,
at least in a range of large satellite masses. 
\end{abstract}

\bodymatter

\section{The dynamical friction process}
\label{dfp}
The deceleration on a test mass $M$ at position identified by the radius vector $\mathbf r$ exerted by the \lq sea\rq~ of particles of mass $m$ where it is moving in, is given by the scattering integral over the whole phase space occupied by the particles \cite{bt} :

\begin{equation}
\frac{d{\mathbf v}_M}{d t} =-{{2m}\over {M+m}}\int_{{\mathbf b}}\int_{{\mathbf v}_m} 
{f(\mathbf{r+b,v}_m)\over b}\frac{{\mathbf v}_M-{\mathbf v}_m}{1+\frac{b^2
|{\mathbf v}_m-{\mathbf v}_M|^4}{G^2(M+m)^2}}{|{\mathbf v}_M-{\mathbf v}_m|}d^3\mathbf{v_m}d^3(\mathbf{r+b}, 
\label{sc_int}
\end{equation}
where $\mathbf b$ is the impact parameter of the $M$--$m$ two-body encounter.
To evaluate this complicated integral along the $M$ test particle path, the so called local approximation formula  for dynamical friction (df) is widely used \cite{Chandra} : 

\begin{equation}
\frac{d\textbf v_M}{dt}=-4\pi^2 G^2  (m+M) \rho(r)F(r,v_M) \ln\Lambda \frac{\textbf{v}_M}{v_M^3},
\label{loc}
\end{equation}

where $F(r,v_M)$ is the fraction of scatterers slower than the test particle and $\ln\Lambda$ is the Coulomb's logarithm.
This expression gives good results also in central galactic cores, but fails when the test particles moves towards the center of a galaxy host whose density profile shows a central cusp, as demonstrated by 
\cite{vic07}. Actually, the central density divergence implies a significant overestimate of the df effect on objects transiting regions close to the center.
To avoid this problem, Arca-Sedda and Capuzzo-Dolcetta \cite{arc12}, in the hypothesis of spherical symmetry for the host galaxy, give an estimate of df by mean of the interpolation formula

\begin{equation}
	\frac{d\textbf{v}_M}{dt}=p(r)\left(\frac{d\textbf{v}_M}{dt}\right)_{cen}+(1-p(r))\left(\frac{d\textbf{v}_M}{dt}\right)_{loc},
	\label{interp}
\end{equation}
where the $cen$ and $loc$ decelerations are obtained via the numerical evaluation of the integral in Eq. \ref{sc_int} and via the usual local approximation, respectively. In Eq.\ref{interp} the weight function, $p(r)$ is assumed as $p(r)=exp(-r/r_c)$, where $r_c$ is a proper scale length.

A more extended and deeper discussion of the treatment of dynamical friction in cuspy galaxies and of the validation of the method described above is found in \cite{arc13}. In the following we just give a short summary of main aspects of the problem and some relevant results.

\section{Massive objects decay in cuspy elliptical galaxies}

To test the approach presented above, we performed a set of $N$-body simulations of the motion of massive objects in galaxy models assumed as \cite{deh93} spherical density profiles, characterized by a length ($r_s$) and a mass ($M_s$) scale, and diverging as $\rho(r)\propto r^{-\gamma}$ in the inner region.

The simulations were done at varying: i) the satellite mass, 
\\$M/M_s=(5\times 10^{-4},10^{-3},5\times 10^{-3})$, ii) the $\gamma$ value, $\gamma=(1/2,1,3/2)$, iii) the initial position $(0.2\le r_0/r_s \le 2)$, and, iv) the type of orbit. The $N$--body simulations were performed by using HiGPUs, a $6^{th}$ order Hermite integrator running on parallel CPU+GPU systems, developed by \cite{cd13}. 
On the other side, the {\it one body} problem of motion of the satellite of mass $M$ in the Dehnen's potential with the semianalytic treatment of dynamical friction described in Sect. \ref{dfp} was integrated by mean of a $6^{th}$ order Runge-Kutta-Nystr\"om method. 

The comparison of the direct $N$--body and semianalytic results over a set of test cases show that the difference in evaluating the decay time is less than $1\%$ for radial orbits and below $10\%$ for circular orbits. This indicates as possible and reliable to simulate the satellite orbital decay by mean of the semyanalitic approach avoiding the use of direct $N$--body simulations, which would require a significant amount of computational time when covering large inetervals of the allowed initial conditions and parameters.

The most important output is the decay time ($t_{df}$), that is the time the object of mass $M$ requires to reduce its motion to a small oscillation around the center of the host galaxy. The local approximation formula leads to an inverse mass dependence for the decay time, $t_{df}\propto M^{-1}$.

The analysis of our results, however, shows that the decay time has a somewhat different scaling with the satellite mass:
\begin{equation}
t_{df}\propto M^{-0.66}.
\end{equation}

\section{Results and applications}

\begin{figure}
\centering
\includegraphics[scale=0.5]{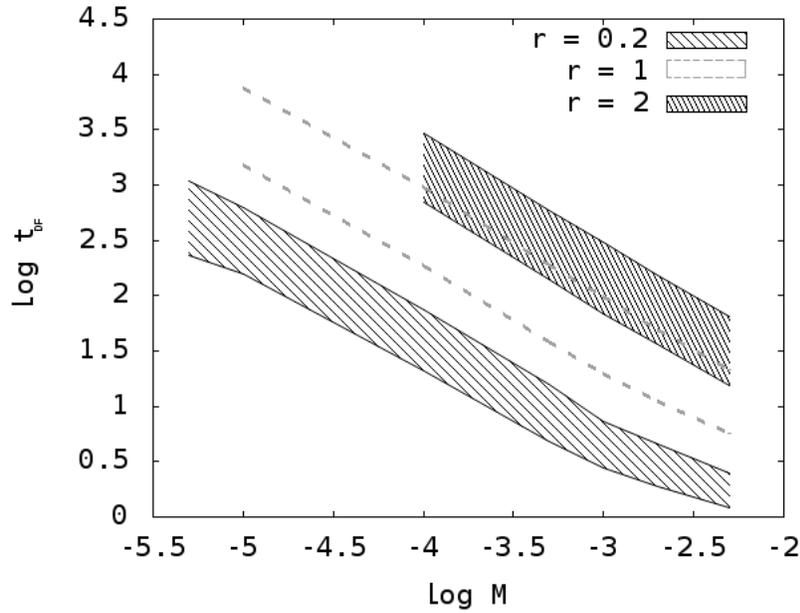}
\caption{\textit{Dynamical friction decay time (in units of crossing time $\tau$) as function of the satellite mass. The three different areas in the plot 
are delimited by a lower boundary (radial orbits) and an upper (circular orbits) and correspond to test masses start moving at the initial distance, 
$r_0$ in units of $r_s$, as labelled at top right. All values refer to the Dehnen's $\gamma=1$ model.}}
\label{grphDF}
\end{figure}

Just to show some preliminary results, we show in Fig.\ref{grphDF} the df time for circular and radial orbits starting at the same position, $r_0/r_s = 0.2-2$, for several masses decaying in a galaxy model of given  mass $M_s$. The satellite mass range  covers the GC mass range up to dwarf spheroidals (to take into account minor mergers). The results shown in Fig. \ref{grphDF} are easily scaled to physical time units
given the relation expressing crossing time $\tau= r_s^{3/2}(GM_s)^{-1/2}$. 

As an example showing the relevance of dynamical friction in cupy galaxies, let us consider a somewhat standard galaxy as having $M_s=5\times 10^{11}$ M$_\odot$ and $r_s=1$ Kpc. This leads to the age of the Universe (assumed 13 Gyr) to be, in logarithm, $4.3$ in the units of Fig. \ref{grphDF}, i.e. in the uppermost part of that Figure.  

\bibliographystyle{ws-procs975x65}
\bibliography{bibliografia}

\begin{thebibliography}{1}

\bibitem{bt}
J.~Binney and S.~Tremaine, {\em Galactic dynamics} (Princeton University Press,
  Princeton, NJ, USA, 2008).

\bibitem{Chandra}
S.~Chandrasekhar, {\em ApJ.} {\bf 97}, p. 255 (1943).

\bibitem{vic07}
A.~{Vicari}, R.~{Capuzzo-Dolcetta} and D.~{Merritt}, {\em ApJ} {\bf 662}, 797
  (2007).

\bibitem{arc12}
M.~Arca-Sedda and R.~Capuzzo-Dolcetta, Dynamical friction in cuspy galaxies, in
  {\em Advances in computational astrophysics: methods, tools and outcomes\/},
  (Cefal\'u, Italy, 2012).

\bibitem{arc13}
M.~Arca-Sedda and R.~Capuzzo-Dolcetta, {The role of dynamical friction in
  cuspidal galaxies}, in preparation, (2013).

\bibitem{deh93}
W.~{Dehnen}, {\em Mon. Not. Roy. Astron. Soc.} {\bf 265}, p. 250 (1993).

\bibitem{cd13}
R.~{Capuzzo-Dolcetta}, M.~{Spera} and D.~{Punzo}, {\em Journ. Comp. Phys.} {\bf
  236}, 580 (2013).

\end{thebibliography}
\end{document}